\documentclass[aps, prl, twocolumn, showpacs, superscriptaddress]{revtex4}

\usepackage{graphicx} 
\usepackage{dcolumn} 
\usepackage{bm} 
\usepackage{amsmath}
\usepackage{SIunits}
\usepackage{times}
\usepackage[ansinew]{inputenc}
\usepackage{upgreek}

\begin{document}

\title{Quantum Brownian motion at strong dissipation probed by superconducting tunnel junctions}

\author{Berthold J\"ack}
\email[Corresponding author; electronic address:\ ]{berthold.jack@alumni.epfl.ch}
\thanks{Present address:\ Princeton University, Joseph Henry Laboratory at the Department of Physics, Princeton, NJ 08544, USA}
\affiliation{Max-Planck-Institut f\"ur Festk\"orperforschung, 70569 Stuttgart, Germany}
\author{Jacob Senkpiel}
\affiliation{Max-Planck-Institut f\"ur Festk\"orperforschung, 70569 Stuttgart, Germany}
\author{Markus Etzkorn}
\affiliation{Max-Planck-Institut f\"ur Festk\"orperforschung, 70569 Stuttgart, Germany}
\author{Joachim Ankerhold}
\affiliation{Institut f\"{u}r Komplexe Quantensysteme and IQST, Universit\"{a}t Ulm, 89069 Ulm, Germany}
\author{Christian R. Ast}
\affiliation{Max-Planck-Institut f\"ur Festk\"orperforschung, 70569 Stuttgart, Germany}
\author{Klaus Kern}
\affiliation{Max-Planck-Institut f\"ur Festk\"orperforschung, 70569 Stuttgart, Germany}
\affiliation{Institut de Physique de la Mati{\`e}re Condens{\'e}e, Ecole Polytechnique F{\'e}d{\'e}rale de Lausanne, 1015 Lausanne, Switzerland}
\date{\today}


\begin{abstract}
We have studied the temporal evolution of a quantum system subjected to strong dissipation at ultra-low temperatures where the system-bath interaction represents the leading energy scale. In this regime, theory predicts the time evolution of the system to follow a generalization of the classical Smoluchowski description, the quantum Smoluchowski equation, thus, exhibiting quantum Brownian motion characteristics. For this purpose, we have investigated the phase dynamics of a superconducting tunnel junction in the presence of high damping. We performed current-biased measurements on the small-capacitance Josephson junction of a scanning tunneling microscope placed in a low impedance environment at milli-Kelvin temperatures. We can describe our experimental findings by a quantum diffusion model with high accuracy in agreement with theoretical predications based on the quantum Smoluchowski equation. In this way we experimentally demonstrate that quantum systems subjected to strong dissipation follow quasi-classical dynamics with significant quantum effects as the leading corrections.
\end{abstract}

\maketitle

{\em Introduction.--} Brownian motion -- that is the fate of a heavy particle immersed in a fluid of lighter
particles -- is {\em the} prototype of a dissipative system coupled to a thermal bath \cite{Risken_1984}. Its quantum mechanical analogue can be found in open quantum systems, which have received considerable attention in the last decade \cite{Schoelkopf_2008}. This is mainly due to the experimental progress in fabricating quantum devices on ever growing scales with the intention to control their quantum properties to an unprecedented accuracy. Efforts have thus focused to tame the impact of decoherence and noise in order to preserve fragile features such as entanglement as possible resources for technological applications \cite{Devoret_2013}.

The regime, where dissipation cannot be seen as a perturbation, but tends to completely dominate the system dynamics has received much less attention. This is in sharp contrast to classical non-equilibrium dynamics, where the so-called overdamped regime, also known as the classical Smoluchowski regime (cSM), plays a pivotal role for diffusion phenomena in a broad variety of realizations \cite{Risken_1984}. Theoretically, according to Smoluchowski, this domain is characterized by a separation of time scales between the relaxation of momentum (fast) and the relaxation of position (much slower) implying that on a coarsely grained time scale the latter constitutes the only relevant degree of freedom. The situation in quantum mechanics is more subtle though. Position and momentum are bound together by Heisenberg's uncertainty relation, which is detrimental to the tendency of strong dissipation to induce localization and even dissipative phase transitions \cite{Schmid_1983, Weiss_1999}.

Roughly speaking, a dissipative quantum system is characterized by three typical energy scales, namely, an excitation energy $\hbar\omega_{\rm 0}$, where $\omega_{\rm 0}$ denotes some specific energy scale of the bare system and $\hbar$ the reduced Planck's constant, a coupling energy to the environment $\hbar\gamma$, where $\gamma$ denotes the coupling parameter, and the thermal energy $k_{\rm B}T$ with $k_{\rm B}$ as Boltzmann's constant and $T$ as the temperature. While the realm of classical physics is then defined by the relation $\hbar\gamma, \hbar\omega_0\ll k_{\rm B} T$, the predominantly explored quantum domain of weak system-bath interaction obeys $\hbar\gamma\ll k_{\rm B}T\ll \hbar\omega_{\rm0}$ with the bare level spacing exceeding all other energy scales. This is the generic situation for cavity and circuit quantum electrodynamical set-ups \cite{Schoelkopf_2008}.

\begin{figure}[h!]
\centering
\includegraphics[width=1\columnwidth]{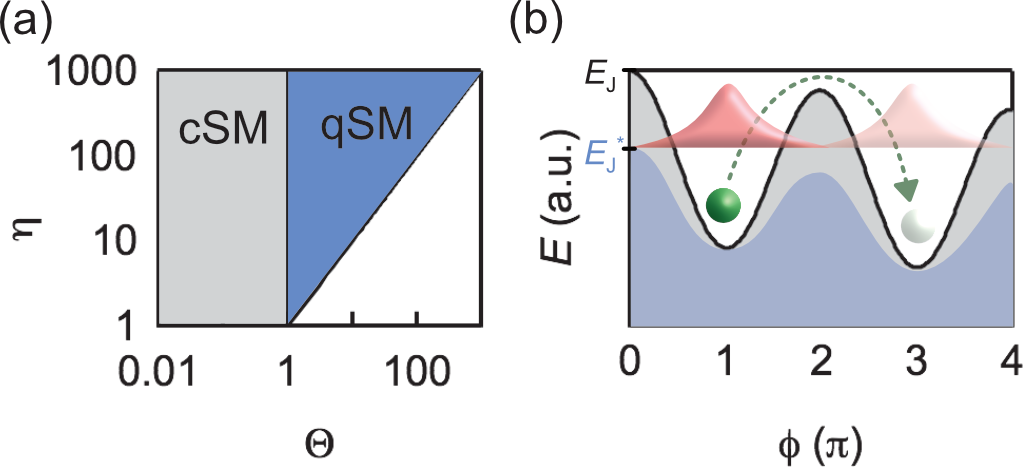}
\caption{(a) Parameter diagram for the dynamics of overdamped phase diffusion in a superconducting tunnel junction. The regime of overdamped quantum diffusion (qSM, blue) emerges at lower temperatures from the classical Smoluchowski domain (cSM, grey) when the dimensionless friction $\eta$ sufficiently exceeds the dimensionless inverse temperature $\Theta$. (b) The phase dynamics corresponds to the dissipative quantum dynamics of a particle in a washboard potential (grey solid), which appears as macroscopic quantum tunneling for weak friction, i.\ e. the tunneling of the phase wavefunction through the potential wall (red wavepackage), and as quantum diffusion in the qSM domain, for which quantum fluctuations effectively reduce the barrier height (blue solid). Classical phase diffusion is illustrated as the thermally activated escape of a particle (green dot) over the potential barrier.}
\label{fig_1}
\end{figure}

The quantum range of strong dissipation is complementary, i.\ e.\ $k_{\rm B}T\ll \hbar\omega_{\rm0} \ll \hbar\gamma$. It has been predicted by theory that in this domain a separation of time scales and thus a quantum Smoluchowski regime (qSM) exists indeed, cf.~Fig.\,\ref{fig_1}(a) \cite{Ankerhold_2001}. Quantum Brownian motion in the qSM is almost classical, however, substantially influenced by quantum fluctuations yielding quantum diffusion characteristics. Consequently, in all processes sensitive to these fluctuations the dynamics is predicted to deviate strongly from the classical one. Hence, to explore the qSM is not only of fundamental interest, but also of direct relevance for strongly condensed phase systems, yet, its experimental observation has been elusive so far.

The aim of this Letter is to close this gap. For this purpose, we study the dynamics of the phase $\phi$ as a continuous collective degree of freedom in a superconducting tunnel junction in the qSM regime. In fact, in the past superconducting circuits have proven to serve as ideal testbeds to explore quantum dissipative phenomena as e.g.\ macroscopic quantum tunneling (MQT), transitions from quantum to classical, or dephasing and decoherence \cite{Simmonds_1971, Leggett_1981, Devoret_1985, Bertet_2005, Ithier_2005}. Here, we access the hitherto untouched qSM domain by investigating the current-voltage characteristics (IVC) of current biased small capacitance tunnel junctions in an ultra-low temperature scanning tunneling microscope (STM) \cite{Assig_2013, Ast_2015}. Its phase dynamics is equivalent to quantum diffusion along a tilted washboard potential under strong damping, cf.\,Fig.\,\ref{fig_1}(b), referring to a low impedance environment with ohmic-resistance $R_{\rm DC}$ much smaller than the quantum resistance $R_{\rm Q}=h/4e^2$, i.e.\ $\rho\equiv R_{\rm DC}/R_{\rm Q}\ll 1$  ($e$ denotes the elementary charge) \cite{Grabert_1998, Ankerhold_2004}. Accessing the qSM regime where quantum diffusion can be observed further requires that the dimensionless friction $\eta=E_{\rm C}/(2\pi^2 \rho^2 E_{\rm J})$, with charging energy $E_{\rm C}$, tunnel coupling $E_{\rm J}$, and dimensionless impedance $\rho$ sufficiently exceeds the dimensionless inverse temperature $\Theta=\beta E_{\rm C}/(2\pi^2 \rho)$ with $\beta=1/k_{\rm B}T$. In terms of experimentally accessible circuit parameters, this condition corresponds to $E_{\rm C}\gg E_{\rm J}$ for a low impedance environment $\rho\ll1$ and milli-Kelvin temperatures \cite{Jaeck_2016}.

\begin{figure}[h!]
\centering
\includegraphics[width=1\columnwidth]{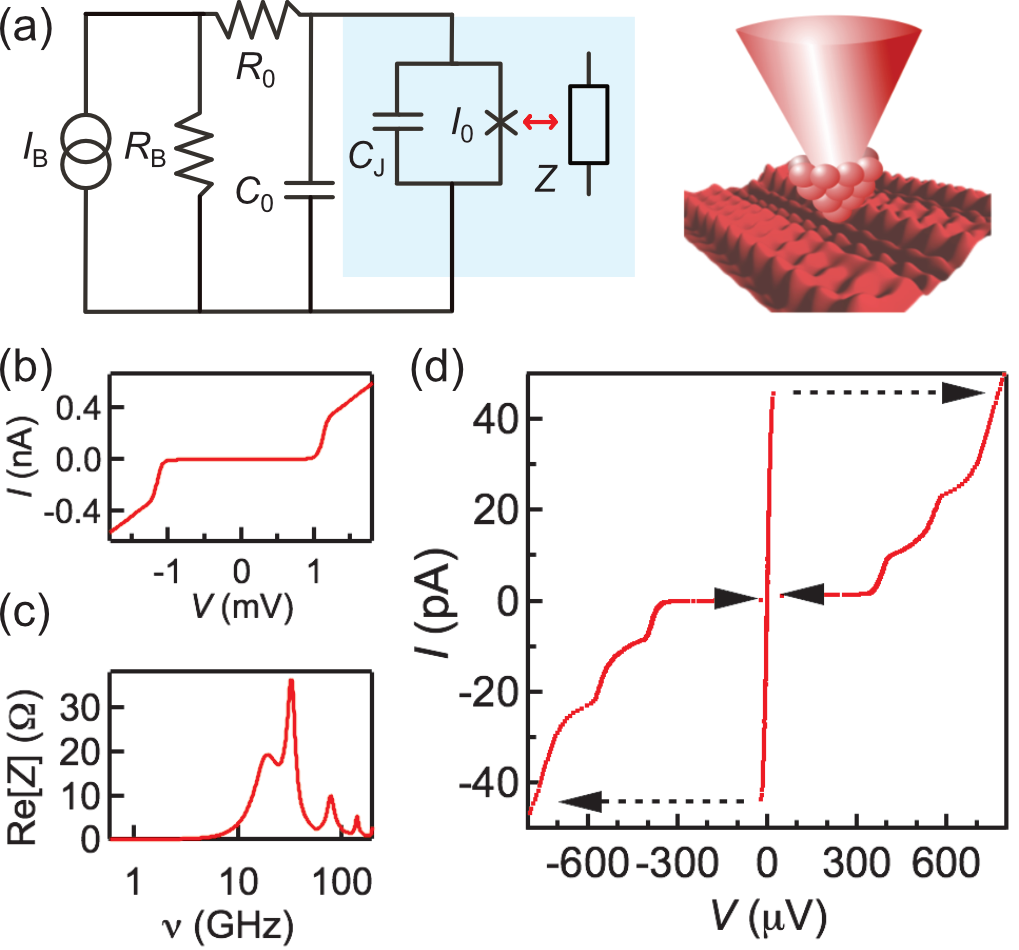}
\caption{(a) Left side: Circuit diagram of the experimental setup with $I_{\text{B}}$ as the bias current source, $R_{\text{B}}$ as the source impedance, $R_{\text{0}}$ and $C_{\text{0}}$ as the load-line resistor and shunt capacitor, respectively. The junction and its direct environment is highlighted by the blue box, where $I_{\text{0}}$ and $C_{\text{J}}$ denote the junction element and capacitor, respectively and $Z$ denotes the environmental impedance. Right side: Schematic representation of the STM tip on top of the reconstructed V(100) surface \cite{Jaeck_2016}. (b) IVC of the superconducting STM tunnel junction measured at $0.004\,G_{\text{0}}$. (c) Simulated real part $\Re[Z]$ of the frequency-dependent environmental impedance \cite{Jaeck_2015}. (d) In-gap region of an IVC from a current-biased measurement at $G_{\text{N}}=0.074\,G_{\text{0}}$.}
\label{fig_2}
\end{figure}

  {\em Theory.--}The supercurrent through a superconducting tunnel junction is determined by its phase dynamics according to the first Josephson relation, $I_{\rm J} = I_{\rm 0} \sin(\phi)$ with the critical current $I_{\rm 0}= 2 e E_{\rm J}/\hbar$. In the qSM regime and for a pure DC environment the diffusion of the phase $\phi$ occurs in a washboard potential tilted by a bias current $I_{\rm B}$ as depicted in Fig.\,\ref{fig_1}(b) \cite{Grabert_1998, Ankerhold_2004}. Below the switching current $I_{\rm S}$, the phase diffuses from one well to another due to the interplay of thermal and quantum fluctuations, this way acquiring a finite velocity $\dot{\phi}\neq0$, cf.\,\ref{fig_1}(b). According to Josephson's second relation, $\langle \dot{\phi}\rangle=(\hbar/2e) V$, this velocity is related to a measurable voltage drop across the junction.
The corresponding Cooper pair current then takes the compact form
  \begin{equation}\label{IZ}
  I_J^{\rm qSM}= \frac{e\rho\beta\pi}{\hbar}\,  (E_{\rm J}^*)^2\, \frac{\beta e V}{(\beta e V)^2+\pi^2 \rho^2}
  \end{equation}
  with a renormalized Josephson energy
  \begin{equation}\label{IZ2}
  E_{\rm J}^*= E_{\rm J}\, \rho^{\rho}\left(\frac{\beta E_{\rm C}}{2\pi^2}\right)^{-\rho} {\rm e}^{-\rho c_{\rm 0}}\, ,
  \end{equation}
where $c_{\rm 0}=0.5772\ldots$ denotes Euler's constant \cite{Grabert_1998, Ankerhold_2004}. It turns out that the above expression can also be understood as the quantum generalization of the corresponding classical Smoluchowski-type of treatment for thermal phase diffusion, the Ivancheko Z'ilberman approach \cite{Zilberman_1969}, by replacing the bare tunnel coupling with its renormalized value   $E_{\rm J}^*$. The explicit dependence of $E_{\rm J}^*$ on $E_{\rm C}$ in Eq.\,\ref{IZ2} reflects the impact of charge fluctuations on the phase dynamics and thus, in the mechanical analogue, the presence of momentum fluctuations in the overdamped diffusion of position. The simultaneous interplay of classical and quantum diffusion leading to Eq.\,\ref{IZ}, therefore, corresponds to quantum Brownian motion dynamics. The physical interpretation is that quantum fluctuations of the phase close to the top of the washboard potential barrier depicted in Fig.\,\ref{fig_1}(b), effectively reduce its height considerably \cite{Ankerhold_2004}. We note, that one may see this effect as complementary to MQT, occurring at low temperatures and weak damping in the opposite domain $E_{\text{J}}\gg E_{\text{C}}$, \cite{Devoret_1985}. As has been demonstrated theoretically, quantum Brownian motion dynamics substantially reduces the maximum supercurrent $I_{\text{S}}$, in the following referred to as switching current, much below its critical current $I_{\text{0}}$ \cite{Grabert_1998, Ankerhold_2004}. We will employ this effect as an experimental probe for the detection of overdamped quantum phase diffusion and to distinguish it from its classical counterpart. In the following, we will provide convincing experimental data for the validity of Eq.\,\ref{IZ} in the qSM regime.

  {\em Superconducting circuit.--} The circuit diagram of the experimental setup is depicted in Fig.\,\ref{fig_2}(a). The atomic-scale tunnel junction appears between a superconducting vanadium STM tip and a vanadium (100) single crystal surface, having a small junction capacitance $C_{\text{J}}$ of a few
femtofarads \cite{Jaeck_2016}. The junction and its direct electromagnetic environment are thermalized at the base temperature of our dilution refrigerator STM of $T=15\,$mK \cite{Assig_2013}, so that one easily arrives at $\Theta\gg1$ (see Fig.\,\ref{fig_1}(a)). A typical IVC measured over a broad voltage range at a low normal state conductivity $G_{\text{N}}=0.004\,G_{\text{0}}$ with $G_{\text{0}}=1/(2 R_{\text{Q}})$ being the conductance quantum, is shown in Fig.\,\ref{fig_2}(b) and exhibits a well-developed superconducting gap. We determine the normal state conductivity by a linear fit to the metallic part of the spectrum with a relative deviation of $5\,\%$; detailed analysis of the superconducting properties of STM tip and sample, the preparation and measurement procedure is described in great detail in Ref.\,\cite{Jaeck_2016}. Since the set-up is operated in the deep tunneling regime at $G_{\text{N}}\ll G_{\text{0}}$ with $E_{\text{J}}$ directly proportional to $G_{\text{N}}$, one has $E_{\text{J}}\ll E_{\text{C}}$ and the condition $1\ll \Theta\ll \eta$ is essentially always fulfilled.

\begin{figure}[h!]
\centering
\includegraphics[width=1\columnwidth]{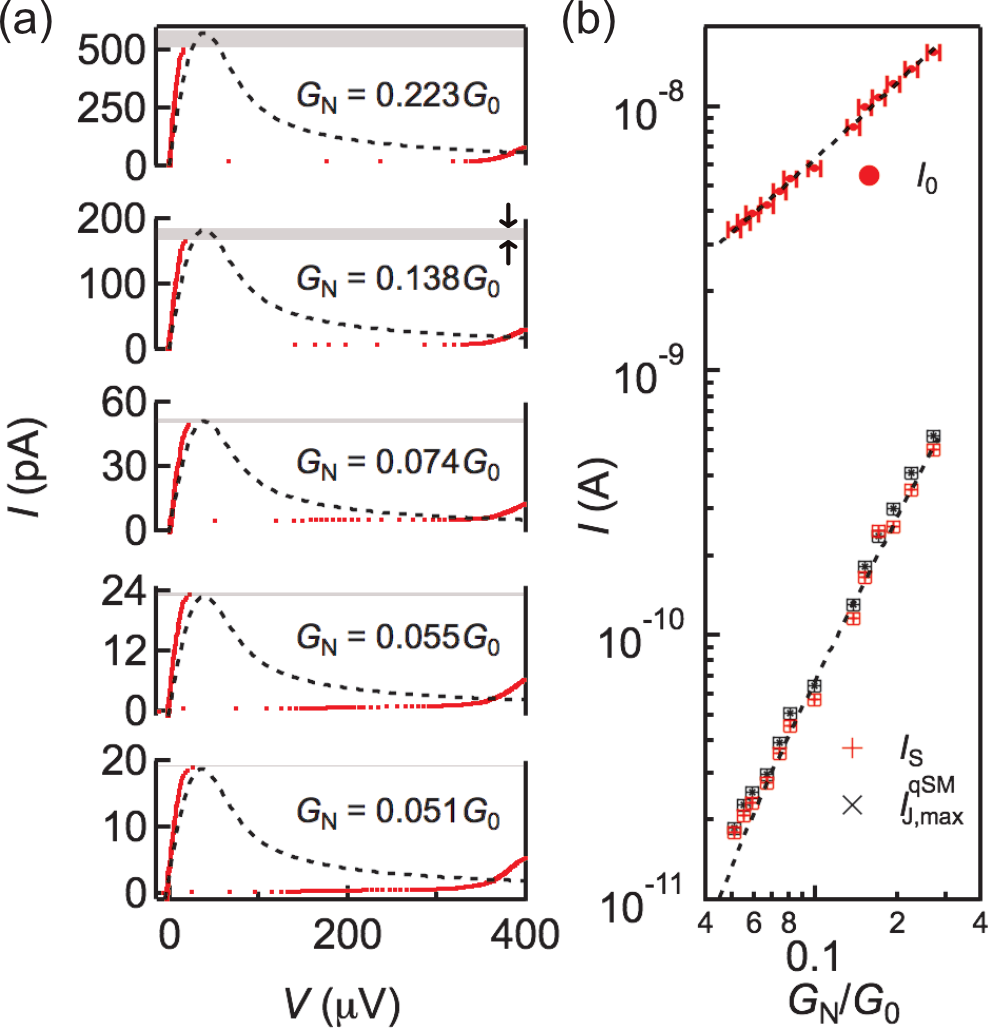}
\caption{(a) Zoom into the low voltage regime of IVCs from current-biased measurements (red dots) at different indicated values of $G_{\text{N}}/G_{\text{0}}$. The dashed lines display the corresponding theoretical IVCs calculated by using the QPD model. The quantitative deviation between experiment and calculation is highlighted via the grey bars for each measurement. (b) Upper part: Calculated critical Josephson current $I_{\text{0}}$ (red dots) and linear fit (black, dashed line) as a function of $G_{\text{N}}/G_{\text{0}}$ (error bars are contained within the symbols). Lower part: Experimental switching current, $I_{\text{S}}$ (red crosses), calculated switching current, $I^{\rm qSM}_{J, \rm max}$ (black crosses), as well as a quadratic fit (black, dashed line) as a function of $G_{\text{N}}/G_{\text{0}}$.}
\label{fig_3}
\end{figure}

In order to minimize the influence of the biasing circuit onto the phase dynamics, we separate the time-scales of the junction phase and the biasing circuit by using a large shunt capacitor $C_{\text{0}}= 3\,$nF and a load-line resistor of $R_{\text{0}}= 3.5\,$k$\Omega$ \cite{Joyez_1999}. Owing to the large output impedance of our constant current source, $R_{\text{B}}=1.33\,$G$\Omega$, the circuit features a horizontal load-line. Concerning the electromagnetic environment of the tunnel junction, we obtain a virtual DC impedance for frequencies up to the low GHz regime by choosing an STM tip of adequate length, moving the tip resonance modes in the environmental impedance $Z(\nu)$ to higher frequencies \cite{Jaeck_2015}. This is illustrated by the frequency-dependent part of the simulated system impedance in Fig.\,\ref{fig_2}(c), where the tip resonance modes appear as sharp peaks \cite{Jaeck_2015}. In the low frequency limit $\nu\to 0$, the environmental impedance $Z(0)$ is, therefore, dominated by the coupling of the tunneling Cooper pairs to the electromagnetic vacuum in the gap between tip and sample (tunnel barrier) with the STM being operated in ultra-high vacuum conditions. Thus, the vacuum impedance $R_{\rm DC}=Z(\nu\to 0)=377\,\Omega$ determines the DC impedance in the vicinity of the tunnel junction, whereas the resistance of the leads (transmission lines) is negligibly small. In this way, we realize the required low impedance environment, $\rho\ll 1$ inducing the required high damping to the phase dynamics and the quantum phase diffusion model of Eq.\,\ref{IZ} becomes applicable to our experiment at small voltages around zero bias \cite{Grabert_1994}.

{\em Results.--} We have measured current-biased IVCs at a total of 14 different values of the normal state conductivity from $0.05\,G_{\text{0}}<G_{\text{N}}<0.3\,G_{\text{0}}$ by changing the vacuum gap width between STM tip and the sample surface. In Fig.\,\ref{fig_2}(d), we plot one example measured at $G_{\text{N}}=0.073\,G_{\text{0}}$ to illustrate the general properties of our experimental setup. For increasing bias current starting from zero, the current follows a phase-diffusion branch at small voltages $V\neq0$, before switching to a dissipative in-gap current. Due to the horizontal load-line of our circuit, we can only access regions of positive differential conductance, yielding a hysteresis in the IVC as shown in Fig.\,\ref{fig_2}(d). For decreasing bias currents starting from $I_{\rm B}>I_{\rm S}$, the current is fixed to the in-gap current before switching back into the phase-diffusion branch. At our experimental conditions, the in-gap current originates from quasiparticle excitations due to life-time effects of Cooper pairs in the STM tip and tunneling of individual Cooper pairs via interaction with resonance modes in the environment \cite{Jaeck_2015, Eltschka_2014, Eltschka_2015}.

Considering the experimental current amplitude of the data shown in Fig.\,\ref{fig_2}(d), we find the switching current $I_{\rm S}=49\pm1\,$pA reduced by approximately two orders of magnitude in comparison to the calculated critical current $I_{\text{0}}=4.80\pm0.24\,$nA \cite{Ambegaokar_1963, Jaeck_2016}. Thermal energy being comparatively small $E_{\text{J}}\beta\geq5$ (see Refs.\,\cite{Jaeck_2016, SI} how to determine the $E_{\text{J}}$ values), classical phase diffusion cannot be the dominant process here. Instead, this strong reduction of $I_{\rm S}$ indicates a major relevance of quantum fluctuations. To better understand this observation, we will in the following present a series of IVCs measured at different values of $G_{\text{N}}/G_{\text{0}}$ and quantitatively analyze these curves by applying the quantum diffusion model of Eq.\,1. We emphasize that varying the normal state conductivity at fixed temperature provides us with an ideal handle to tune the $\eta\propto E_{\text{J}}^{-1} \propto G_{\text{N}}^{-1}$ value along a vertical axis in the phase diagram in Fig.\,\ref{fig_1}(a).

In Fig.\,\ref{fig_3}(a), we present a series of IVCs measured at different values of $G_{\text{N}}/G_{\text{0}}$ and take a closer look at the low voltage regime $V<30\,\upmu$V for positive bias currents. Towards higher normal state conductivities, we observe a strong and non-linear increase of the switching current $I_{\text{S}}$. This observation is illustrated in Fig.\,\ref{fig_3}(b), where we plot the extracted $I_{\text{S}}$ values as a function of $G_{\text{N}}/G_{\text{0}}$ and fit a quadratic dependence $I_{\text{S}}\propto G_{\text{N}}^{2}$, as it is predicted by Eq.\,\ref{IZ}. We additionally find $I_{\text{S}}$ strongly reduced in comparison to the calculated critical Josephson current $I_{\text{0}}\propto G_{\text{N}}$ also shown in Fig.\,\ref{fig_3}(b) for all measurements, as we have mentioned before. In combination, these findings reveal the major relevance of quantum fluctuations in the limit $E_{\text{J}}\ll E_{\text{C}}$ \cite{Ingold_1992, SI}. 

To quantitatively analyze the experimental IVCs in Fig.\,\ref{fig_3}(a), we calculate the corresponding theoretical IVCs by using Eq.\,\ref{IZ} for each $G_{\text{N}}/G_{\text{0}}$ value and plot them alongside the experimental data. We obtain the necessary experimental values of $E_{\text{C}}$, $E_{\text{J}}$ and $T$ by performing additional voltage-biased measurements on the same tunnel junction at the same $G_{\text{N}}/G_{\text{0}}$ values. Fitting the resulting IVCs with $P(E)$-theory allows us to extract these parameters with high precision (see Refs.\,\cite{Jaeck_2016} and \cite{SI} for more details) \cite{Devoret_1990, Averin_1990}. Owing to the small junction capacitance in the absence of a large shunt capacitor in the junction vicinity, thermally induced capacitive noise broaden the measured IVCs. We account for this effect by convolving the calculated IVC from Eq.\,\ref{IZ} with a normalized Gaussian function $P_{\text{N}}(E)$ of width $\sigma=\sqrt{2E_{\text{C}}/\beta}$ \cite{Jaeck_2016, Ast_2015}. 

Comparing experimental and theoretical IVCs in Fig.\,\ref{fig_3}(a), we find excellent quantitative agreement in the low voltage regime $V<30\,\upmu$V for small normal state conductivity values. We recall that our experimental setup can only access regions of positive differential resistance. At these small voltages, we attribute the slightly different slope of the phase diffusion branch between experiment and theory to small deviations of our environmental impedance from a perfect DC behavior as required for the derivation of Eq.\,\ref{IZ}. By contrast, considering the IVCs measured at high $G_{\text{N}}$ values shown in Fig\,\ref{fig_3}(a), we observe significant quantitative deviations in the current amplitude between theory and experiment. We will analyze these deviations in the following in more detail and also compare our experimental data with a classical diffusion model.

\begin{figure}[h!]
\centering
\includegraphics[width=1\columnwidth]{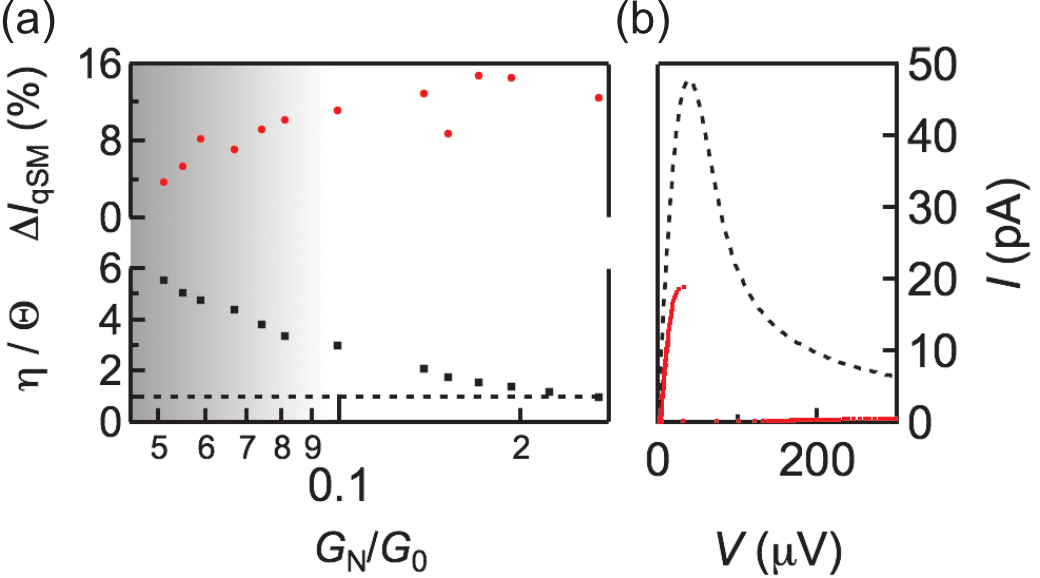}
\caption{(a) Relative deviation between theoretical and experimental current maxima $\Delta I_{\rm qSM}$ as well as the qSM regime boundary $\eta/\Theta$ as a function of the normalized normal state conductivity. (b) Comparison between experimental data (red, solid) and an IVC calculated (black, dashed) from the purely classical IZ model at a normal state conductivity of $G_{\text{N}}=0.051\,G_{\text{0}}$.}
\label{fig_4}
\end{figure}

{\em Discussion.--} We start our discussion by comparing the maximum of the measured IVC, $I_{\rm S}$, with the maximum of the calculated IVC, $I^{\rm qSM}_{J, \rm max}$, as a measure for the agreement between experiment and theory. In Fig.\,\ref{fig_4}(a), we plot their relative deviation $\Delta I_{\rm qSM} =|I^{\rm qSM}_{J, \rm max}-I_{\rm S}|/I^{\rm qSM}_{J, \rm max}$ as a function of $G_{\text{N}}/G_{\text{0}}$. Furthermore, we include the conditions for the qSM regime, $\eta \gg \Theta \gg 1$ by using the same set of parameters from the $P(E)$-fits as before. We obtain $\Theta=122$ for all measurements and plot the dependence of $\eta/\Theta$ on $G_{\text{N}}/G_{\text{0}}$ in Fig.\,\ref{fig_4}(a). In the low conductivity regime, the condition for the observation of qSM dynamics is fulfilled $\eta/\Theta>4$, explaining the very small deviations between theory and experiment $3\,\%<\Delta I_{\rm qSM}<9\,\%$. By contrast, we find $\Delta I_{\rm qSM}>12\,\%$ at high conductivities where the condition for qSM dynamics is violated $\eta/\Theta\approx1$, yielding an overall consistent picture. 

Additionally, we compare our experimental data with the {\em classical} IZ model (IZ) in which phase diffusion only occurs via thermally activated escape over the washboard potential barrier displaying classical Brownian motion dynamics, cf.\ Fig.\,\ref{fig_1}(b) \cite{Ambegaokar_1969, Zilberman_1969}. We plot an experimental IVC together with a theoretical IVC calculated using the IZ model and the same parameter set from $P(E)$-theory as before for a normal state conductivity of $G_{\text{N}}=0.051\,G_{\text{0}}$ in Fig.\,\ref{fig_4}(b). As can be seen, theory based on classical diffusion cannot reproduce our experimental data and largely overestimates the IVC amplitude. This observation clearly underlines the high relevance of quantum fluctuations, significantly reducing the washboard potential barrier height and, thus the switching current in comparison to classical phase diffusion dynamics. Formally, this effect yields the renormalized Josephson coupling energy $E_{\text{J}}^*$ in Eq.\,\ref{IZ2}, as compared for the bare $E_{\text{J}}$ in the classical regime.

Together, the accurate agreement between experiment and theory at low conductivities demonstrates that the overdamped quantum phase dynamics of our tunnel junction can be described in the framework of the qSM equation \cite{Ankerhold_2004}. At high conductivities, however, the condition for a separation of time scales of position and momentum dynamics, $\eta\gg \Theta$, is violated and, accordingly, the qSM equation is not applicable anymore. Hence, our experimental study reveals that the time evolution of an overdamped quantum system follows quasi-classical dynamics with significant quantum-mechanical corrections in leading order that is quantum Brownian motion \cite{Ankerhold_2001}. Based on its fundamental character in the framework of quantum statistics, our study should be of general relevance for the fields of superconducting quantum circuits \cite{Devoret_2013}, quantum gases \cite{Bloch_2008} and nano-mechanical oscillators \cite{Kippenberg_2014}, for instance.

{\em Conclusion.--} We have studied the temporal evolution of a quantum system subjected to strong dissipation in the framework of the qSM equation. We experimentally investigated the phase dynamics of a small-capacitance Josephson junction in an STM placed in a low-impedance environment at ultra-low temperatures by means of current-biased measurements. We can theoretically describe the measured IVCs in the low voltage regime by using a quantum phase diffusion model derived from the qSM equation with high accuracy. In this way, our study reveals that the dynamics of an overdamped quantum system corresponds to quantum Brownian motion, which is quasi-classical dynamics with significant quantum-mechanical corrections. In addition, our study demonstrates the unique potential of ultra-low temperature STM to address questions in the field of mesoscopic transport and quantum statistics. We envision that future experiments on the temperature dependent phase dynamics in superconducting tunnel junctions should reveal the transition from quantum-mechanical to classical Smoluchowski dynamics.

It is our pleasure to acknowledge inspiring discussions with P.~W. Anderson, D. Esteve, J. Pekola, F. Tafuri, M. Ternes, S. Kochen and A. Yazdani.

\end{document}